\documentclass[apj]{emulateapj}

\usepackage{graphicx,float}

\def\cm{{\rm\thinspace cm}}

\def\erg{{\rm\thinspace erg}}

\def\km{{\rm\thinspace km}}

\def\Mpc{{\rm\thinspace Mpc}}

\def\s{{\rm\thinspace s}}

\def\ergps{\hbox{$\erg\s^{-1}\,$}}

\def\kmpspMpc{\hbox{$\km\s^{-1}\Mpc^{-1}\,$}}

\def\pcmsq{\hbox{$\cm^{-2}\,$}}

\shorttitle{\sc{No bar Influence on Average BH Growth}}
\shortauthors{{\sc A.D.~Goulding et al.}}

\begin{document}


\title{Galaxy-scale Bars in Late-type Sloan Digital Sky Survey
  Galaxies Do Not Influence the Average Accretion Rates of
  Supermassive Black Holes}


\author{A. D. Goulding\altaffilmark{1}$^{,\dagger}$,
  E. Matthaey\altaffilmark{1}, J. E. Greene\altaffilmark{1},
  R. C. Hickox\altaffilmark{2}, D. M. Alexander\altaffilmark{3},
  W. R. Forman\altaffilmark{4}, C. Jones\altaffilmark{4},
  B.~D.~Lehmer\altaffilmark{5}, S. Griffis\altaffilmark{2,6},
  S. Kanek\altaffilmark{2} \& M. Oulmakki\altaffilmark{2}}

\affil{$^1$Department of Astrophysical Sciences, Princeton University, Ivy Lane, Princeton, NJ 08544, USA; $^{\dagger}$goulding@astro.princeton.edu}
\affil{$^2$Department of Physics and Astronomy, Dartmouth College, 6127 Wilder Laboratory, Hanover, NH 03755, USA}
\affil{$^3$Center for Extragalactic Astronomy, Department of Physics, University of Durham, South Road, Durham DH1 3LE, UK}
\affil{$^4$Harvard-Smithsonian Center for Astrophysics, 60 Garden Street, Cambridge, MA 02138, USA}
\affil{$^5$Department of Physics, University of Arkansas, 226 Physics Building, 825 West Dickson Street, Fayetteville, AR 72701, USA}
\affil{$^6$Department of Computer Science, Harvard University, 33 Oxford St, Cambridge, MA 02138, USA}

\begin{abstract}
  Galaxy-scale bars are expected to provide an effective means for
  driving material towards the central region in spiral galaxies, and
  possibly feeding supermassive black holes (BHs). Here we
  present a statistically-complete study of the effect of bars on
  average BH accretion. From a well-selected sample of 50,794 spiral
  galaxies (with $M_* \sim 0.2-30 \times 10^{10} M_{\odot}$) extracted
  from the Sloan Digital Sky Survey Galaxy Zoo 2 project, we separate
  those sources considered to contain galaxy-scale bars from those
  that do not. Using archival data taken by the {\it Chandra} X-ray
  Observatory, we identify X-ray luminous
  ($L_X \gtrsim 10^{41} \ergps$) active galactic nuclei (AGN) and
  perform an X-ray stacking analysis on the remaining X-ray undetected
  sources. Through X-ray stacking, we derive a time-averaged look at
  accretion for galaxies at fixed stellar mass and star formation
  rate, finding that the average nuclear accretion rates of galaxies
  with bar structures are fully consistent with those lacking bars
  ($\dot{M}_{\rm acc} \approx 3 \times
  10^{-5}$~M$_{\odot}$~yr$^{-1}$).
  Hence, we robustly conclude that large-scale bars have little or no
  effect on the average growth of BHs in nearby ($z<0.15$) galaxies
  over gigayear timescales.
\end{abstract}


\section{Introduction} \label{sec:intro}

The seminal discovery that the masses of supermassive black holes
(BHs) show a proportional relationship with their stellar spheroids
(e.g., \citealt{Magorrian:1998aa,Gebhardt:2000aa,ferrarese00}) implies
a strong physical association between the growth of BHs and the
evolution of their host galaxies (e.g.,
\citealt{croton06,hopkins06}). With the advent of wide-field
extragalactic surveys, we now have a panchromatic view of millions of
galaxies and active galactic nuclei (AGN). Indeed, studies performed
using data from these multiwavelength surveys have revolutionized our
understanding of galaxy formation, and the connections between
galaxies and their central BHs (e.g.,
\citealt{juneau13,Daddi:2007aa,Hickox:2009aa,donley12,Goulding:2014aa,Jones:2016aa};
for a review see \citealt{dma12}).

The evolution of luminous AGN activity ($L_X \gtrsim 10^{44} \ergps$)
has now been traced to $z \approx 5$ (e.g.,
\citealt{ueda03,dma03a,hasinger05,Mortlock:2011aa,Xue:2011aa,Vito:2013aa,Vito:2014aa,Aird:2015aa,Georgakakis:2015aa,Brandt:2015aa,Vito:2016aa,Marchesi:2016aa}). These
deep and wide-field studies have revealed that galaxies undergoing a
rapid stage of evolution are often found to host powerful and
extremely luminous AGN. Such luminous AGN activity is frequently
associated with recent galaxy--galaxy interactions
(\citealt{Ellison:2011aa,Cisternas:2011aa,Silverman:2011aa,Treister:2012aa,Glikman:2015aa,Hong:2015aa}). Though
still an active area of discussion, the majority of more moderate
luminosity AGN ($L_X \sim 10^{42}-10^{44} \ergps$) at $z\sim 0.1-2$
appear to be hosted in disk-like systems, that do not appear to have
recently undergone a significant merger (e.g.,
\citealt{Schawinski:2011aa,Kocevski:2012aa,Simmons:2012aa,Rosario:2013aa,Cisternas:2015aa,Cheung:2015aa}). Hence,
the evolution of these galaxies is relatively slow and largely
decoupled from the cosmic framework; in turn, they must be growing
their BHs through more secular processes \citep[see
below;][]{hopkins06c,schawinski12}.

While gas-rich mergers can readily provide a plentiful supply of
material to accrete onto a BH, the dominant mechanism through which
BHs grow in secularly-evolving (e.g., \citealt{Kormendy:2004aa})
spiral galaxies is still poorly understood. Secular evolution is
expected to be primarily driven through large-scale instabilities,
such as dynamic bars (e.g.,
\citealt{Shlosman:1993aa,Courteau:1996aa,Bournaud:2002aa,Sheth:2005aa,Cheung:2013aa,Sanchez-Janssen:2013aa,Sellwood:2014aa}),
which may have been induced by an early merger (Noguchi 1987). Through
angular momentum exchange, bars generate radial flows that are capable
of transporting kiloparsec-scale gas down to approximately parsec
scales, close to the galaxy center (e.g.,
\citealt{Shlosman:1989aa,Friedli:1993aa,Wang:2012aa,Fanali:2015aa}). The
resulting cool gas reservoir may serve to feed the BH, and hence, bars
are proposed as viable mechanisms to trigger AGN activity in spiral
galaxies (e.g.,
\citealt{Shlosman:1990ab,Wada:1995aa,Bournaud:2002aa,Athanassoula:2003aa,Laurikainen:2004aa,Jogee:2006aa}).

However, to date, no definitive connection has been made between
large-scale bar structures in spiral galaxies and the presence of a
central moderate-luminosity AGN. Previous studies have primarily
focused on searching for an enhancement in the bar fraction of AGN,
typically indentified through optical emission line diagnostics, over
control samples of spiral galaxies within focused high-quality nearby
and large format statistical studies in the optical and near-infrared
regimes, often with seemingly contradictory results (for a recent
review of these results, see Table 1 of
\citealt{Galloway:2015aa}). More indirectly, there has also been
speculation that barred spirals are offset (by a factor $\sim$2) from
the standard BH--bulge relation
(\citealt{Graham:2009aa,Brown:2013aa}); though others have found
little difference between barred and non-barred systems (e.g., Xiao et
al. 2011).

Previous studies have often been hampered by small number
statistics. However, Galloway et al. (2015) recently harnessed the
enormous wealth of morphological information available in the second
release of the Sloan Digital Sky Survey citizen science project,
Galaxy Zoo, to find that, at fixed stellar mass and rest-frame $u - r$
color, galaxies hosting AGN show a $\sim$16\% increase in their bar
fraction. Seemingly contradictory results arise from the citizen
science project `Galaxy Zoo: Hubble' \citep{Cheung:2015aa}, finding no
evidence for bar enhancement in AGN. It is therefore unclear whether a
bar is the dominant factor for growing a BH in the absense of a recent
merger.

A possible scenario that could explain the existence of AGN in only
some barred galaxies is that AGN activity is a stochastic
process. Simulations find that the physical processes that feed BH
growth on small spatial scales are unlikely to be smooth or
continuous, leading BH accretion to vary dramatically on short
(megayear) timescales (\citealt{Hopkins:2010ad,Hickox:2014aa}). This
variation may hide a strong underlying correlation with longer-lived
galaxy properties, such as star-formation
(e.g. \citealt{Gabor:2013aa,Hickox:2014aa,Thacker:2014aa,Delvecchio:2015aa})
or the existence of a large-scale bar, which can remain in-situ for
significant fractions of a gigayear. A statistically complete study
focused on assessing average AGN activity as a function of bar
properties (rather than assessing bars in AGN hosts), and hence, a
time-averaged look at BH accretion, may serve to alleviate current
tensions in the literature.

X-ray observations provide a robust detection of AGN activity that is
less affected by obscuration than optical light. But, through the use
of `X-ray stacking techniques', these observations may also provide
aggregate measurements of BH accretion rate for suitably-selected
galaxy samples. In relatively small ($< 10$~deg$^2$) contiguous
fields, X-ray stacking has been used effectively to search for faint
X-ray signals of heavily-obscured or low-luminosity accretion (e.g.,
\citealt{Brandt:2001aa,Hickox:2007aa,Vito:2016aa}). Indeed, when
combining the unprecedented angular resolution of the {\it Chandra}
X-ray Observatory with complementary multi-wavelength data, a number
of studies have revealed significant populations of formally
undetected AGN (e.g.,
\citealt{Daddi:2007aa,Fiore:2008aa,Georgakakis:2008aa,dma11,chen13}),
as well as the average AGN space densities and luminosity functions
across Cosmic time (e.g.,
\citealt{worsley05,Hickox:2006aa,Xue:2011aa}). Most recently, in the
COSMOS field, X-ray stacking has been used, in conjunction with the
{\it Hubble} Space Telescope, as a first step towards addressing the
question of the effectiveness of large-scale bars in driving the
growth of BHs in the absence of a major merger
\citep{Cisternas:2015aa}. Based on their X-ray stacking of barred
versus non-barred galaxies, \cite{Cisternas:2015aa} find that at
$z \sim 0.3-0.8$ the average AGN luminosity is not influenced by the
presence of a bar.

Building on these previous studies, here we combine the unprecendented
wealth of morphological information available from the Sloan Digital
Sky Survey with the inherent statistical power of an X-ray stacking
analysis performed using data from the {\it Chandra} X-ray Observatory
to measure the average BH accretion rates of galaxies that do and do
not contain large-scale bars. In Section 2, we describe the selection
process for samples of barred and non-barred galaxies extracted from
the Sloan Digital Sky Survey and Galaxy Zoo, as well as our X-ray data
and stacking analyses. In Section 3, we ascertain the effect of bars
on the growth of BHs in nearby galaxies, finding no difference in the
average accretion rate of AGN determined from X-ray stacking in barred
galaxies compared with control samples of non-barred systems. In
Section 4, we discuss the implications of our findings and present our
concluding remarks. Throughout this manuscript we adopt a flat
$\Lambda$CDM cosmology with $H_0 = 71 \kmpspMpc$ and $\Omega_M = 0.3$.

\section{Sample Selection} \label{sec:sample}

\begin{figure}[t]
\centering
\includegraphics[width=0.99\linewidth]{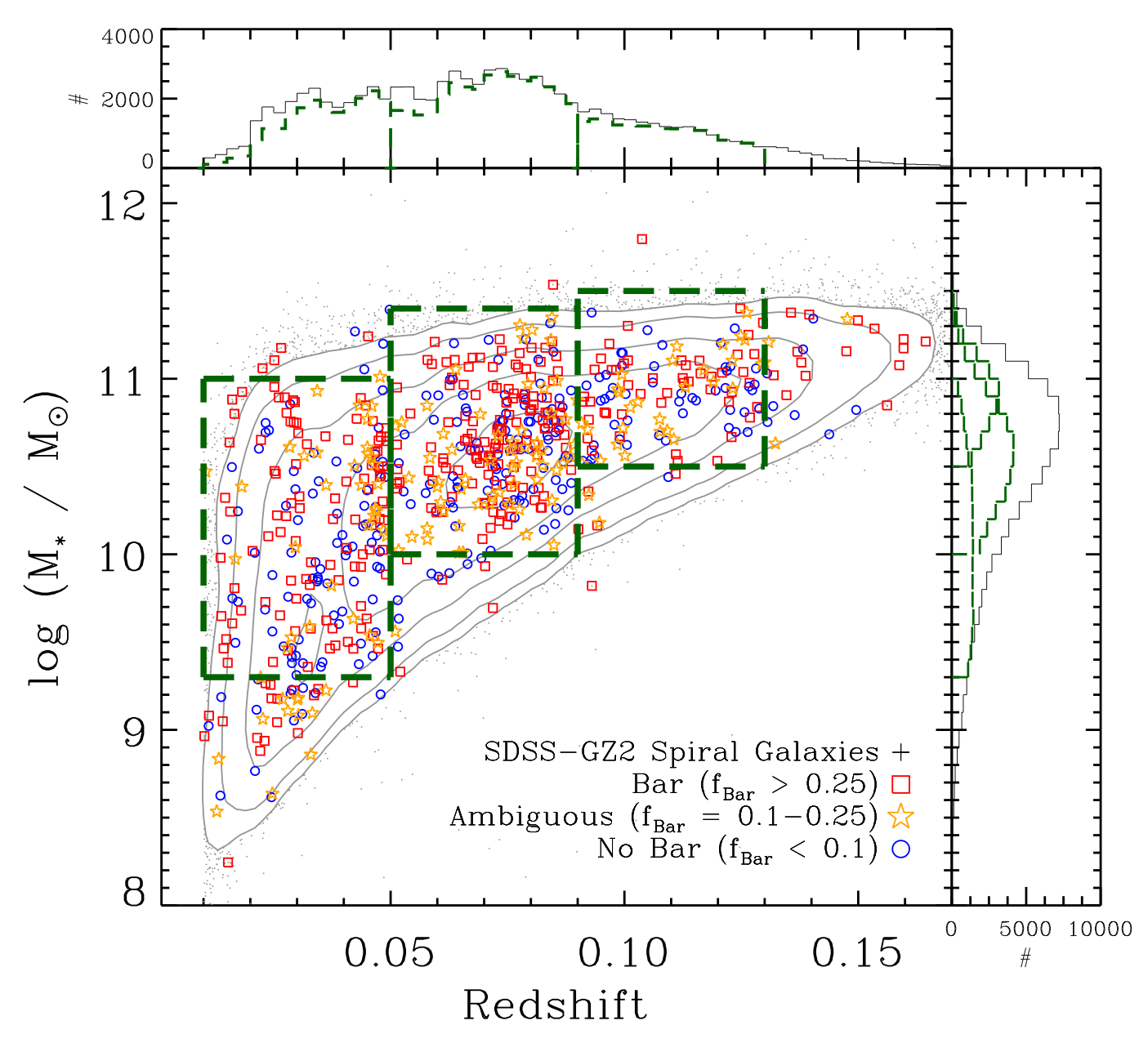}
\vspace{-0.1cm}
\caption{{\bf Main panel:} Stellar mass versus spectroscopic redshift
  for non edge-on ($f^{\rm not}_{\rm Edge-on} > 0.5$) SDSS-GZ2 disk
  galaxies (gray contours). Based on the debiased vote fraction,
  galaxies in the SDSS-GZ2 and {\it Chandra} X-ray ACIS overlap region
  that are determined to have bar structures ($f_{\rm bar} > 0.25$),
  to have no discernable bar ($f_{\rm bar} < 0.1$), and those galaxies
  with ambiguous evidence for a bar ($f_{\rm bar} = 0.1-0.25$) are
  shown with red squares, blue circles and yellow stars,
  respectively. Bounding regions for three selected $M_*-z$ bins used
  throughout \S3 are shown with green-dashed lines. {\bf Upper panel:}
  Redshift distributions for the full SDSS-GZ2 sample (gray) and the
  galaxies selected in the dashed boxes in the main panel (green
  dash). {\bf Right Panel} Stellar mass distributions for the full
  SDSS-GZ2 sample (gray), and the three $M_*-z$-dependent selections
  shown in the main panel (green dash).}
\label{fig:mass_z}
\end{figure}

In this section, we describe the selection of our uniform and
optical-property matched nearby spiral galaxy sample. The source
sample is selected from the $\sim 120$~deg$^2$ overlap region between
the Sloan Digital Sky Survey and serendipitous observations performed
up to and including Cycle 16 with the Advanced CCD Imaging
Spectrometer (ACIS) onboard the {\it Chandra} X-ray
Observatory. Additionally, we outline our X-ray stacking methodology
used to measure the average X-ray luminosities of this galaxy sample.

\subsection{Barred galaxies in SDSS Galaxy-Zoo 2}

We begin by constructing a parent sample of spiral galaxies from which
to select our sub-samples of galaxies with and without large-scale
bars. Our sample of local galaxies is selected from the seventh data
release of the Sloan Digital Sky Survey (hereafter, SDSS-DR7;
\citealt{Abazajian:2009aa}), where we select all spectroscopically
targeted $r<17.77$~magnitude galaxies in the redshift range
$z \sim 0.01-0.17$. Following \cite{Galloway:2015aa}, the lower
redshift limit excludes those galaxies whose angular size
significantly exceeds the size of the spectroscopic fiber.

Morphological cuts on the SDSS parent sample are made using
visual-classification data gathered from the citizen science project,
Galaxy Zoo (\citealt{Lintott:2008aa,Lintott:2011aa}). Specifically, we
harness the second release of the Galaxy Zoo project (hereafter,
SDSS-GZ2) that provides detailed morphological statistics on
$\sim$300,000 SDSS galaxies, including inclination angle, existence of
spiral arms, bulge dominance and galactic bars. These statistics are a
collation of the responses to a set of hierarchical decision-tree
questions, posed to citizen scientists, regarding morphological
features that may or may not be present in 3-color images of SDSS
galaxies (for details see \citealt{Willett:2013aa}).

We identified a sample of disk galaxies by selecting all systems in
SDSS-GZ2 at $0.01 \lesssim z \lesssim 0.17$ that had been visually
classified by at least 20 people (i.e., the galaxy had at least 20
responses to the zeroth node question: `Is the galaxy simply smooth
and rounded, with no sign of a disk?'), and where the fraction of
classifiers identifying disk features (i.e., positive responses to the
zeroth node question) was $f_{\rm disk} \geq 0.227$. Large-scale bars
may be difficult to identify in objects that are highly inclined,
hence, we remove likely edge-on systems with the classification
fraction threshold $f^{\rm not}_{\rm Edge-on} > 0.5$. Selecting
against edge-on systems will further remove systematic bias against
obscured AGN where the AGN emission is being extinguished by
line-of-sight material residing in the host galaxy (e.g.,
\citealt{GA09,lagos11,goulding12a}), and not necessarily a small-scale
dust/gas-rich torus. Performing our stacking analyses at X-ray
energies will then allow us to mitigate the effects of the small-scale
obscurer. Our chosen cuts on the SDSS-GZ2 are similar to those
suggested by \cite{Willett:2013aa} from their in-depth analysis of the
entire SDSS-GZ2 data and catalog. After our zeroth order morphological
cuts, the SDSS-GZ2 sample contains a robust sample of 96,767 spiral
galaxies.

\begin{figure*}[t]
\centering
\includegraphics[width=0.99\textwidth]{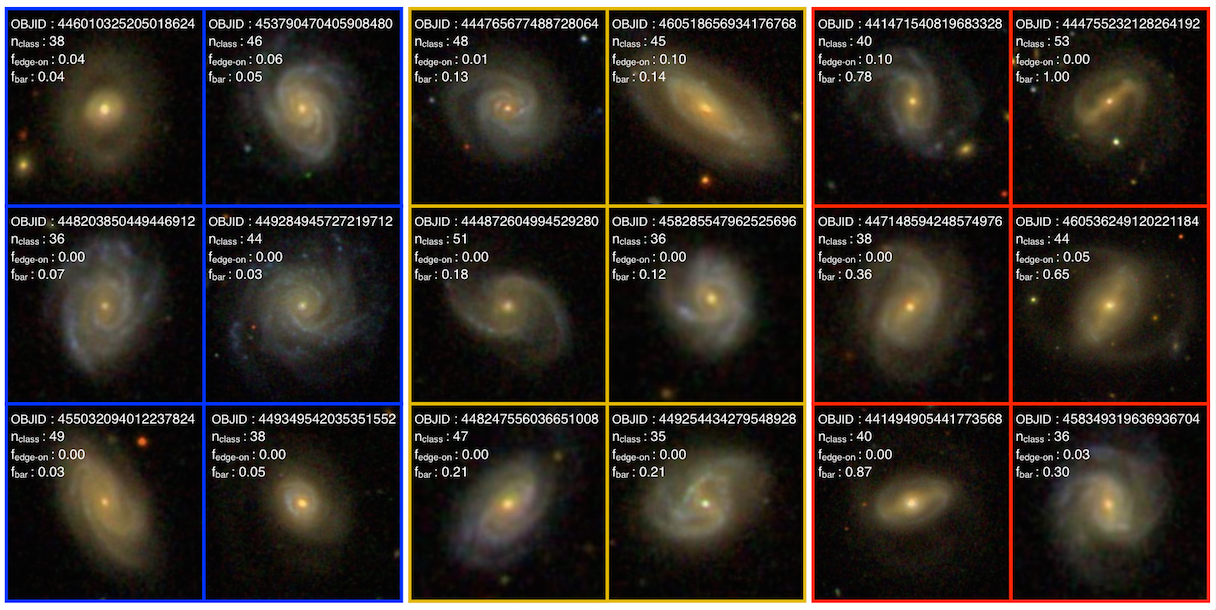}
\vspace{-0.1cm}
\caption{Three color ($gri$) composite SDSS-Legacy images for a random
  set of sources from the SDSS-GZ2 galaxy sample. Colored lines
  highlight sources with $f_{\rm bar} > 0.25$ (red),
  $f_{\rm bar} = 0.1-0.25$ (yellow), and $f_{\rm bar} < 0.1$
  (blue). Individual galaxies are labeled with their SDSS unique
  obejct identifier number (objid), the number of citizen scientists
  who classified the galaxy ($n_{\rm class}$), the edge-on debiased
  vote fraction ($f_{\rm edge-on} = 1 - f^{\rm not}_{\rm edge-on}$),
  and the galaxy-bar present debiased vote fraction ($f_{\rm bar}$).}
\label{fig:galaxies}
\end{figure*}

X-ray emission produced due to AGN activity and/or star forming
processes is known to be a function of galaxy stellar mass (e.g.,
\citealt{Lehmer:2010aa,Lehmer:2016aa}). Hence to avoid bias and
incompleteness, we require our sample to be complete in stellar mass
in a given redshift bin. Due to the $r<17.77$~mag selection, the SDSS
spectroscopic galaxy sample forms a stellar mass complete sample
\citep{brinchmann04}. At the low-redshift limit of the survey,
SDSS-DR7 is complete to $M_* \sim 10^8 M_{\odot}$, and allows
volume-limited galaxy samples that are mass-complete to be constructed
to $z < 0.22$. To obtain estimates of the stellar mass and
star-formation rates for the 96,767 spiral galaxies in our SDSS-GZ2
sample, we use the value-added MPA-JHU spectroscopic catalog of
\cite{brinchmann04}. \cite{brinchmann04} used the SDSS {\tt cmodel}
photometry and fiber spectroscopy to fit stellar population synthesis
templates using a Bayesian methodology to derive the physical
properties. Here we use the 50th percentile of the log total stellar
mass PDF (i.e., the median estimates of the total stellar masses), the
star-formation rate within the fiber, and the scaled total
star-formation rate. Median uncertainties are 0.09~dex and 0.14~dex on
$M_*$ and SFR, respectively.

In Figure~\ref{fig:mass_z} we show the distribution of the
$\sim$10$^5$ SDSS-GZ2 spirals with stellar mass estimates in the
MPA-JHU catalog. We identify three redshift bins, 0.01--0.05,
0.05-0.09 and 0.09-0.13, where our spiral sample is complete to a
given stellar mass threshold (green dashed boxes in Fig.~1). These
three $M_* - z$ sub-samples form the basis for our X-ray analyses
presented in section~\ref{sec:stackfast}.

We distinguish between galaxies with and without bars using the
debiased fraction of votes\footnote{The debiased fraction of votes
  are the fraction of votes cast by citizen scientists on a particular
  classification question, which have been weighted for consistency
  and adjusted for redshift dependent classification bias. See Section
  3.3 of \cite{Willett:2013aa} for further details.}  measured from
the SDSS-GZ2 question: `is there a sign of a bar feature through the
center of the galaxy?'  (hereafter, $f_{\rm bar}$). We select only
spiral galaxies in our $M_* - z$ samples where the number of votes on
the `bar question' was $\geq 5$. The average number of `bar question'
votes per galaxy for the sample is 13. A naive cut of
$f_{\rm bar} > 0.5$ provides a bar fraction of only $\sim 18$\%. This
bar fraction is substantially below the known fraction of local
($D<40$~Mpc) galaxies that contain a bar signature ($\gtrsim 59$\%)
based on near-IR identification, which is far less susceptible to dust
obscuration effects than optical identifications
(\citealt{Eskridge:2000aa,Menendez-Delmestre:2007aa}).

Furthermore, a simple analysis of the distribution of $f_{\rm bar}$
for SDSS-GZ2 spiral galaxies, shows it to be roughly exponentially
peaked at $f_{\rm bar} \sim 0.0$ with a declining tail towards
$f_{\rm bar} \sim 1.0$. The shape of this distribution is likely
driven by observational biases, as the true probability distribution
of bar versus non-bars in spiral galaxies must be far more
bi-modal. Hence, $f_{\rm bar}$ does not directly map to the
probability of a bar existing in a particular spiral galaxy.

In order to separate barred galaxies from non-barred galaxies we use
the known nearby bar-fraction ($\gtrsim 59$\%), and match this to the
cumulative distribution of $f_{\rm bar}$ for the SDSS-GZ2 parent
spiral sample. A demarcation of $f_{\rm bar} \gtrsim 0.15$ results in
a bar fraction of $\sim 55$\%.  Given small number statistics, and
allowing for the possibility of false-positives, we conservatively use
a threshold of $f_{\rm bar} \geq 0.25$, providing us with a `clean'
bar sample (bar-fraction of $\sim 40$\%).\footnote{Our bar selection
  threshold is also similar to a cut of $f_{\rm bar} > 0.3$ that was
  determined from the independent analysis of
  \cite{Willett:2013aa}. See also Galloway et al. (2015).} Similarly,
if we adopt a threshold of $f_{\rm bar} < 0.1$, we achieve a `clean'
non-barred spiral sample with a non-bar fraction of $\sim 40$\%. The
remaining $\sim 20$\% of galaxies in our $M_* - z$ samples with
$0.1 < f_{\rm bar} < 0.25$, in a purely statistical sense, have
ambiguous evidence for the presence of a bar. This is further
evidenced in Fig~\ref{fig:galaxies}, where we provide example images
used in GalazyZoo2 to classify the galaxies. Sources with
$f_{\rm bar} \geq 0.25$ seemingly have strong and/or obvious bar
structures, while those with $0.1 < f_{\rm bar} < 0.25$ have much
weaker and/or difficult to identify bars in their SDSS 3-color
images. In Table~1, we provide the number of non edge-on spiral
galaxies in our $M_* - z$ samples determined to have bars (28,733
galaxies), no bars (28,728 galaxies), and ambiguous evidence for bars
(15,474 galaxies).

Finally, to statistically compare the AGN properties of our $M_* - z$
samples based on their inferred bar properties, it is important to
ensure that the distributions of the first-order galaxy properties
(such as star-formation rate; stellar mass) are well matched between
the bar/no-bar sub-populations. In Figure~3, we provide the
star-formation rate (SFR) measured within the spectroscopic fiber and
the stellar mass distributions of the three $M_* - z$ samples,
separated based on their inferred bar properties. Based on two-sample
Kolmogorov-Smirnov (K-S) tests, we find evidence at the $>$99.99\% level
that the SFR distributions for the bar and non-bar samples are not
drawn from the same distributions ($P<10^{-7}$; see also
\citealt{Oh:2012aa}). It is further evident from the histograms
presented in Figure~3 that, as a function of $M_*$ and $z$, there is
an increasing fraction of barred galaxies with low SFRs over those
without bars. It is beyond the scope of this study to provide a full
explanation for this phenomenon, though we suggest that it maybe
easier to visually identify bars in systems that lack significant
levels of SF or that star-forming regions may systematically reside at
larger radii in barred galaxies (Robichaud et al. 2017).

For the purposes of our experiment, it is sufficient to simply ensure
that the distributions in galaxy properties are similar for the barred
and non-barred stacked samples. Hence, to mitigate these effects, we
apply SFR cuts of log~SFR/($M_{\odot}$/yr)$> -1.8$, $>-1.4$, and
$> -1.0$ for our $M_* - z$ redshift bins of 0.01--0.05, 0.05--0.09 and
0.09--0.13, respectively. We find that these cuts are sufficient to
simultaneously remove the tensions between the SFR and $M_*$
distributions for the SDSS-GZ2 galaxies included in our X-ray stacking
analysis (described in the next section). Comparing the SFR
distributions for the barred and non-barred that are covered by {\it
  Chandra} X-ray observations, from two-sample K-S tests we find
$P \gtrsim 0.2$ between the $0.01 < z < 0.05$ and $0.05 < z< 0.09$
samples, while we still find some evidence for tension
($P \sim 0.015$) between the samples at $0.09<z<0.13$.

While not fully essential for our analyses, as we also take care to
statistically account for X-ray emission from SF in our stacking
analysis, requiring similar distributions in $M_*$ and SFR is still a
useful endeavor when comparing the X-ray properties of galaxies as it
equally aggregates the contamination to the X-ray emission from SF
processes, while also allowing us to (statistically) select sources
hosting similar mass BHs (through the $M_*-M_{\rm BH}$ correlation).

\begin{figure*}[t]
\centering
\includegraphics[width=0.99\textwidth]{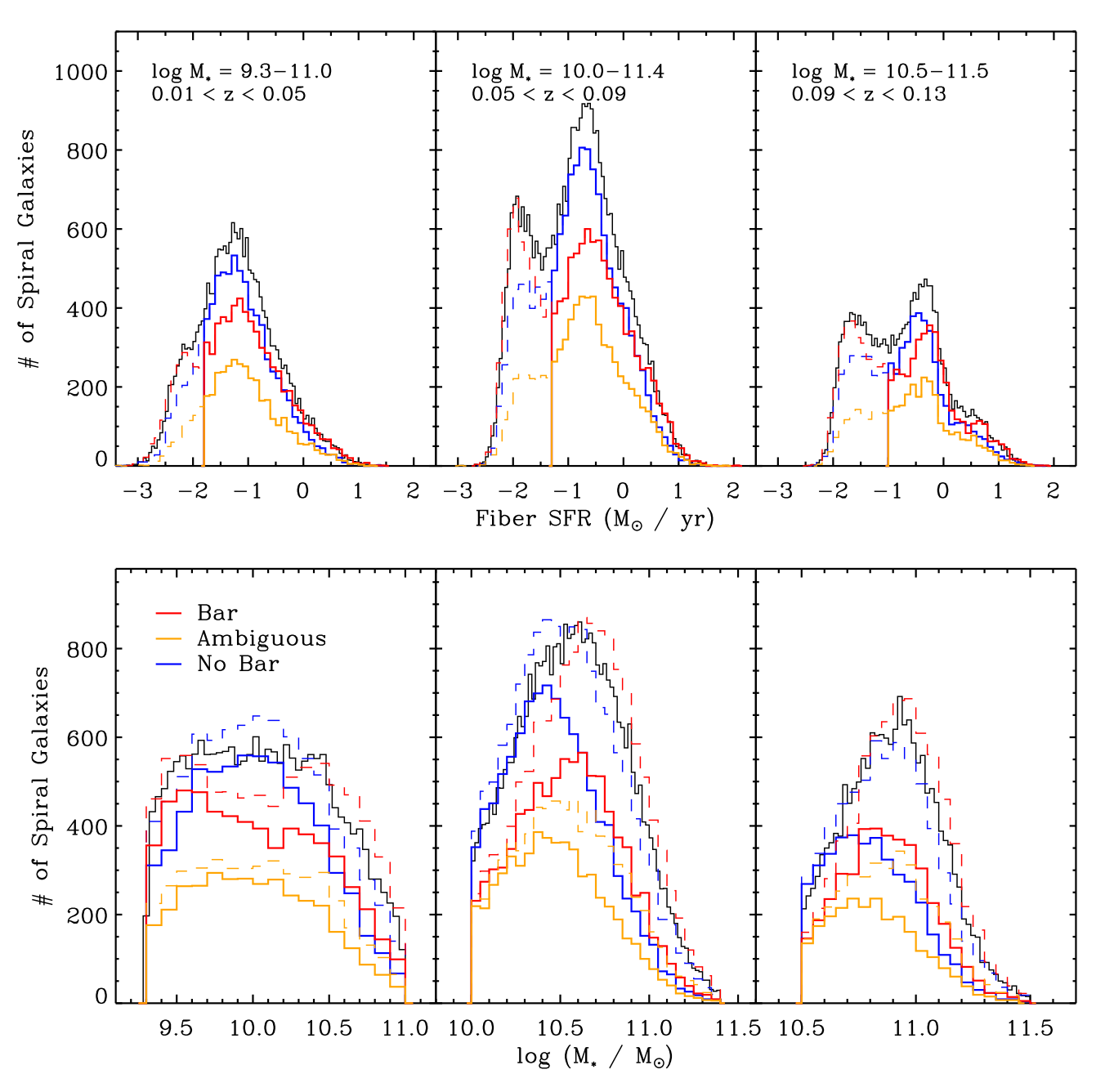}
\vspace{-0.1cm}
\caption{{\bf Upper Panel:} Histograms of star formation rates within
  the SDSS spectroscopic fiber for all SDSS-GZ2 galaxies in the
  $M_*-z$ bins (black solid lines) shown in Fig.~1. {\bf Lower panel:}
  Histograms of total stellar mass. {\bf Common:} Colored lines
  represent SDSS-GZ2 galaxies with $f_{\rm bar} > 0.25$ (red),
  $f_{\rm bar} = 0.1-0.25$ (yellow), and $f_{\rm bar} < 0.1$
  (blue). Dashed lines represent the full galaxy samples in their
  respective bar category. Solid lines represent the galaxies used
  throughout our analyses in Section 3 after matching the samples on
  star-formation rate.}
\label{fig:histos}
\end{figure*}

\begin{table*}[ht!]

\begin{center}
\setlength{\tabcolsep}{1.0mm}
\caption{Galaxy Sample and X-ray Stacking Results}
\begin{tabular}{lccccccccccccccccc}

\hline \hline

\multicolumn{6}{c}{} &
\multicolumn{4}{c}{---------~~~0.5--2 keV~~~---------} &
\multicolumn{4}{c}{---------~~~2--7 keV~~~---------} &
\multicolumn{4}{c}{---------~~~0.5--7 keV~~~---------} \\
\multicolumn{1}{c}{log$M_{*}$} &
\multicolumn{1}{c}{$z$} &
\multicolumn{1}{c}{$f_{\rm bar}$} &
\multicolumn{1}{c}{\#} &
\multicolumn{1}{c}{$\bar{t}_{\rm Exp}$} &
\multicolumn{1}{c}{$\Sigma t_{\rm Exp}$} &
\multicolumn{1}{c}{\#} &
\multicolumn{1}{c}{\#} &
\multicolumn{1}{c}{$\bar{L}_{\rm X,AGN}$} &
\multicolumn{1}{c}{$S/N$} &
\multicolumn{1}{c}{\#} &
\multicolumn{1}{c}{\#} &
\multicolumn{1}{c}{$\bar{L}_{\rm X,AGN}$} &
\multicolumn{1}{c}{$S/N$} &
\multicolumn{1}{c}{\#} &
\multicolumn{1}{c}{\#} &
\multicolumn{1}{c}{$\bar{L}_{\rm X,AGN}$} &
\multicolumn{1}{c}{$S/N$} \\
\multicolumn{1}{c}{} &
\multicolumn{1}{c}{} &
\multicolumn{1}{c}{} &
\multicolumn{1}{c}{SDSS} &
\multicolumn{1}{c}{} &
\multicolumn{1}{c}{} &
\multicolumn{1}{c}{Det.} &
\multicolumn{1}{c}{Stack} &
\multicolumn{1}{c}{$10^{39}$} &
\multicolumn{1}{c}{} &
\multicolumn{1}{c}{Det.} &
\multicolumn{1}{c}{Stack} &
\multicolumn{1}{c}{$10^{39}$} &
\multicolumn{1}{c}{} &
\multicolumn{1}{c}{Det.} &
\multicolumn{1}{c}{Stack} &
\multicolumn{1}{c}{$10^{39}$} &
\multicolumn{1}{c}{} \\
\multicolumn{1}{c}{($M_{\odot}$)} &
\multicolumn{1}{c}{} &
\multicolumn{1}{c}{} &
\multicolumn{1}{c}{} &
\multicolumn{1}{c}{(ks)} &
\multicolumn{1}{c}{(ks)} &
\multicolumn{1}{c}{} &
\multicolumn{1}{c}{} &
\multicolumn{1}{c}{(erg~s$^{-1}$)} &
\multicolumn{1}{c}{} &
\multicolumn{1}{c}{} &
\multicolumn{1}{c}{} &
\multicolumn{1}{c}{(erg~s$^{-1}$)} &
\multicolumn{1}{c}{} &
\multicolumn{1}{c}{} &
\multicolumn{1}{c}{} &
\multicolumn{1}{c}{(erg~s$^{-1}$)} &
\multicolumn{1}{c}{} \\
\multicolumn{1}{c}{(1)} &
\multicolumn{1}{c}{(2)} &
\multicolumn{1}{c}{(3)} &
\multicolumn{1}{c}{(4)} &
\multicolumn{1}{c}{(5)} &
\multicolumn{1}{c}{(6)} &
\multicolumn{1}{c}{(7)} &
\multicolumn{1}{c}{(8)} &
\multicolumn{1}{c}{(9)} &
\multicolumn{1}{c}{(10)} &
\multicolumn{1}{c}{(11)} &
\multicolumn{1}{c}{(12)} &
\multicolumn{1}{c}{(13)} &
\multicolumn{1}{c}{(14)} &
\multicolumn{1}{c}{(15)} &
\multicolumn{1}{c}{(16)} &
\multicolumn{1}{c}{(17)} &
\multicolumn{1}{c}{(18)} \\[1.2ex]

\hline

 9.3--11.0 & 0.01--0.05   & $>$0.25   &  5958 &33.8 & 2905.8 & 17 & 63 & $1.56^{+0.42}_{-0.43}$ & 10.7 & 13 & 67 & $2.88^{+0.67}_{-0.65}$ & 8.2 & 18 & 62 & $3.47^{+0.97}_{-1.04}$ & 13.5 \\
           &              & 0.1--0.25 &  3486 &24.5 & 770.0  & 10 & 21 & $1.02^{+0.76}_{-0.62}$ & 5.4  & 10 & 21 & $1.11^{+1.31}_{-0.90}$ & 3.3  & 10 & 21 & $0.59^{+1.70}_{-0.58}$ & 6.3 \\
           &              & $<$0.1    &  6543 &24.7 & 1444.3 & 17 & 57 & $1.01^{+0.68}_{-0.83}$ & 10.1 & 14 & 60 & $1.63^{+0.80}_{-0.61}$ & 6.9  & 16 & 58 & $0.48^{+2.30}_{-0.35}$ & 12.2 \\[1.2ex]
\hline 
10.0--11.4 & 0.05--0.09   & $>$0.25   &  8644 &17.0 & 1448.5 & 17 & 60 & $3.60^{+1.91}_{-1.65}$ & 7.9  & 13 & 64 & $5.75^{+4.00}_{-4.62}$ & 6.1  & 15 & 62 & $7.66^{+4.79}_{-5.48}$ & 10.0 \\
           &              & 0.1--0.25 &  5637 &22.0 & 1035.6 &  8 & 37 & $4.07^{+1.25}_{-1.23}$ & 7.1  &  7 & 38 & $1.31^{+3.79}_{-2.82}$ & 3.8  & 8  & 37 & $5.33^{+3.45}_{-2.31}$ & 8.0 \\ 
           &              & $<$0.1    & 10059 &22.1 & 1552.2 & 13 & 54 & $2.87^{+1.17}_{-1.04}$ & 6.9  & 12 & 55 & $3.39^{+3.06}_{-2.70}$ & 5.0  & 13 & 54 & $5.04^{+3.22}_{-2.68}$ & 8.5 \\[1.2ex]
\hline
10.5--11.5 & 0.09--0.13   & $>$0.25   &  4074 &27.3 & 1090.4  &  8 & 27 & $8.20^{+3.76}_{-2.78}$ & 6.3  & 6  & 29 & $19.2^{+7.81}_{-5.81}$ & 5.5  & 8  & 27 & $22.1^{+10.7}_{-7.59}$ & 8.3 \\
           &              & 0.1--0.25 &  2401 &35.9 & 970.8  &  3 & 19 & $3.19^{+2.48}_{-1.98}$ & 4.7  & 2  & 20 & $13.2^{+6.18}_{-5.00}$ & 5.1  & 3  & 19 & $11.3^{+6.51}_{-5.50}$ & 7.0 \\
           &              &  $<$0.1   &  3992 &33.5 & 1272.8 &  7 & 29 & $4.16^{+1.89}_{-1.77}$ & 4.6  & 3  & 33 & $17.2^{+10.5}_{-12.8}$ & 5.3  & 6  & 30 & $1.43^{+7.41}_{-8.70}$ & 7.1 \\[1.2ex]
\hline \hline
\end{tabular}
\end{center}
($1$) Ranges of the logarithm of the stellar mass in units of solar masses;
($2$) Redshift ranges;
($3$) Fraction of citizen scientists voting for the presence of a bar in the SDSS image;
($4$) Number of SDSS--GZ2 galaxies in each bin matched for similar distributions in star-formation rates measured within the SDSS spectroscopic fiber;
($5$) Mean exposure time of all galaxies covered by Chandra ACIS-I observations in units of kiloseconds;
($6$) Total exposure time for all galaxies covered by Chandra ACIS-I observations in units of kiloseconds;
($7-10$) Number of X-ray detected AGN in the 0.5--2 keV band, number of non X-ray detected galaxies included in the X-ray stack, star-formation subtracted stacked (average) X-ray luminosity (in units of $10^{39} \ergps$) in the 0.5--2 keV band, signal-to-noise of the X-ray emission in the stack above the stacked local background;
($11-14$) As columns 7--9 in the 2--7 keV band;
($14-18$) As columns 7--9 in the 0.5--7 keV band.

\end{table*}

\subsection{STACKFAST: X-ray Stacking Analyses} \label{sec:stackfast}

Stacking analyses are based on the principle that objects at known
positions, which are not detected individually in imaging at a
particular wavelength, may show a significant flux when the
observations are averaged together. These aggregate measurements are
most accurate when both the multi-wavelength properties of the stacked
sample are previously well defined and source number statistics are
sufficiently high to break through the noise floor, and thus, increase
the signal-to-noise and effective depth of the observations. By
definition, these stacked signals then have the added effect of
averaging over the inherent stochasticity of AGN accretion, which is
present in blind selections of AGN samples.

To perform the X-ray stacking of our sample, we used our custom {\sc
  idl}-based software {\sc stackfast}, which is designed specifically
for {\it Chandra} ACIS data and we briefly describe here (see also
\citealt{hickox07c,Hickox:2007aa,Hickox:2009aa,chen13}). {\sc
  stackfast} begins with two elements: a master catalog of input
sources, and a set of uniformly formatted, reduced, and flare-cleaned
{\it Chandra} ACIS data products derived from individual {\it
  Chandra}-ACIS observations (ObsIDs). Here we include all ACIS ObsIDs
that were publicly available from Cycles 1--16 of the {\it Chandra}
mission. ObsIDs were reduced using version 4.6 of the {\sc ciao}
software package and by applying v4.6.5 of the {\sc caldb} calibration
files. Events were screened using the grade set 0,2,3,4,6, and
$3 \sigma$ background flares were removed using the {\tt lc\_clean}
tool. Aspect histograms were constructed using the {\tt aspecthist}
tool available in the {\sc chav} software package, and convolved with
the ACIS chip-map to generate observation exposure maps. For a
detailed explanation of the reduction and processing procedure see
\cite{goulding12b}.

Sources from our master catalog that lie within the field of view of
an ObsID are identified, and events/photon information (position,
energy, grade, and time) are extracted from a $30'' \times 30''$ box
centered on the input source for each associated ObsID. {\sc
  stackfast} also determines the effective exposure time at the
position of the input source for each ObsID, including the effects of
vignetting at large angles from the pointing axis. The result is an
output file containing a stackable X-ray event list and list of
exposure times for each source in the master catalog, and associated
ObsID.

Once the extraction of the stackable database is performed, the next
step for {\sc stackfast} is coadding a specific subset of sources to
yield average background-subtracted X-ray count rates and fluxes in
user-defined energy bands. A key consideration is the variable
point-spread function (PSF) for {\it Chandra} as a function of
off-axis angle, which we characterize by the 90\% energy-encircled
radius ($r_{90}$)\footnote{$r_{90}$ is calculated based on ACIS-I PSF
  maps that are constructed at 0.5-2 keV, 2-7 keV and 0.5-7
  keV.}. {\sc stackfast} extracts photons within $r_{90}$ and assesses
the background on a source-by-source basis, as variations in the PSF
also affects background estimates. Local backgrounds are measured by
masking high significance ({\tt wavdetect} threshold $10^{-7}$)
sources that may be identified in the $30'' \times 30''$ extraction
boxes, and then rescaling the photon counts, in the remaining
extraction area at radii larger than $1.3 \times r_{90}$, to the
source extraction area (within $r_{90}$). These local background
measurements are co-added, and used to measure the signal-to-noise of
the source stacks.

During the co-adding process, we exclude sources that lie within 5$''$
of the pointing position of a particular ObsID, as these sources may
have been the subject of the proposed observation. The inclusion of
these `proposed' sources would introduce a selection bias into the
final X-ray stack. Furthermore, we also exclude input sources that are
significant X-ray detections, fall on chip-gaps, or are positioned at
large off-axis angles $>6'$ in a particular ObsID, as background
estimation may be inaccurate or severely enhanced due to the large
angular size of an X-ray point source ($\sim 5-7.5''$ at
$E \sim 1.5$--7~keV). The final co-adding procedure is extremely
efficient, and thus, readily allows bootstrap realizations to be
performed to derive count rate uncertainties on the stacking
measurements. In any given X-ray energy band, the final outputs from
{\sc stackfast} are average X-ray count rates within $r_{90}$, average
X-ray backgrounds predicted within $r_{90}$, hardness ratios, and
associated statistical uncertainties from 1000 bootstrap
realizations. X-ray count rates are converted to fluxes assuming a
powerlaw with slope $\Gamma = 1.8$ and a {\it Chandra} ACIS-I response
function from Cycle 9 (an average of the collation of observations
used throughout our stacking).

Of the 50,794 spiral galaxies in the $M_* - z$ sub-samples, 468
($\sim 1$\%) of the sources are within 6$'$ of the pointing position
of at least one {\it Chandra} ACIS-I observation, and also are not on
chip-gaps or close ($<5''$) to the pointing position. There are 101
galaxies which we determine to have significant (3$\sigma$) X-ray
emission above the estimated local background that is close to the
optical position. Each of the X-ray detected sources have centroids
within $0.9''$ of the optical position, with a median offset of
$\delta_{\rm X-O} \sim 0.26''$. This results in a sample of 367 X-ray
undetected galaxies that are included as part of our stacking analysis
in Section 3.

\begin{figure*}[t]
\centering
\includegraphics[width=0.99\textwidth]{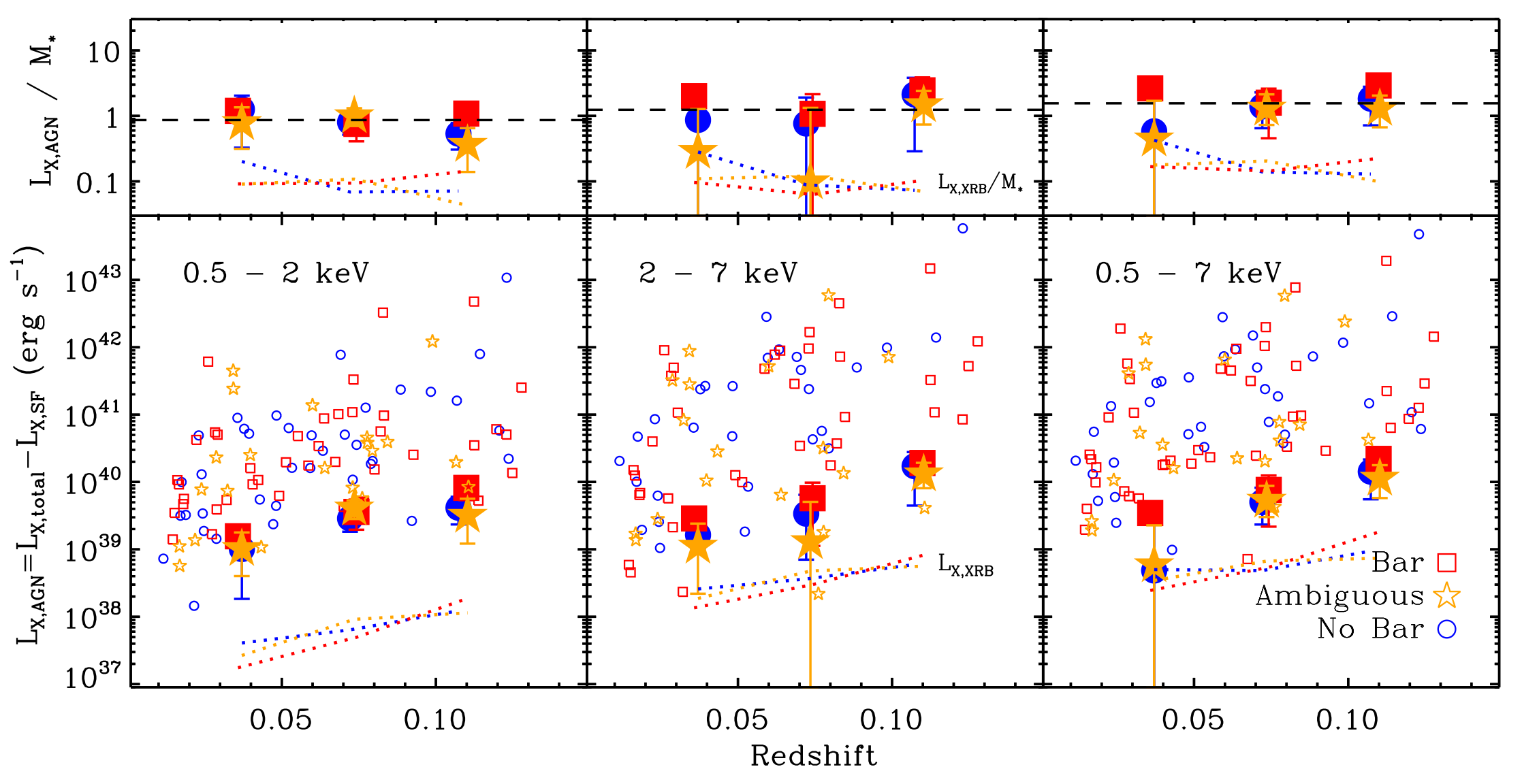}
\caption{{\bf Lower panels:} Star-formation subtracted AGN X-ray
  luminosity versus redshift in the 0.5--2~keV (left column), 2--7~keV
  (center column) and 0.5--7~keV (right column) energy
  bands. Individual redshift bins are matched in $M_*$ and SFR for
  SDSS-GZ2 galaxies in the SDSS-CXO overlap region. SDSS-GZ2 systems
  with $f_{\rm bar} > 0.25$, $f_{\rm bar} = 0.1-0.25$ and
  $f_{\rm bar} < 0.1$ are shown with red squares, blue circles and
  yellow stars, respectively. X-ray detected sources with
  $S/N \gtrsim 5$ close to the optical position are shown with open
  symbols, and luminosities produced from stacking of X-ray undetected
  sources are shown with large filled symbols. 90th percentile
  uncertainties on the X-ray stacks are calculated through Bootstrap
  resampling of the sources within the stacks. For illustration
  purposes, the predicted mean X-ray luminosities produced due to
  stellar processes are shown with dotted lines for each stacked
  redshift bin. {\bf Upper panels:} Specific AGN X-ray luminosities
  (i.e., $L_{\rm X,AGN}$ normalized by median stellar mass in units of
  $10^{29} \ergps M_{\odot}^{-1}$). The thick dash line shows the
  median of the X-ray stacks. We find there is no positive effect
  between the average BH accretion rate and the presence of a
  large-scale bar in nearby spiral galaxies, and when normalized for
  stellar mass this is also independent of redshift.}
\label{fig:lx_z}
\end{figure*}

\section{Results} \label{sec:results}

To search for a connection between the average growth of supermassive
BHs and the existence of a large-scale bar in spiral galaxies, we
perform an X-ray stacking analysis of nearby spiral galaxies. As
described in the previous section, we have selected samples of
SDSS-GZ2 galaxies that are matched in stellar mass, SFR and
redshift. These are separated between barred ($f_{\rm bar} > 0.25$),
non-barred ($f_{\rm bar} < 0.1$) and ambiguously-barred
($f_{\rm bar} = 0.1-0.25$) systems based on the fraction of votes
received from citizen scientists using the SDSS imaging.

\subsection{The average accretion rate of (un)-barred spiral galaxies} \label{subsec:stack}

The majority of BHs residing at galaxy centers are in a low-Eddington
state
($\lambda_{\rm Edd} = L_{\rm AGN}/L_{\rm Edd} \lesssim 10^{-3}$), and
thus, their subsequent AGN emission would fall substantially below the
detection threshold of a typical extragalactic survey. Our primary
goal is to harness the power of X-ray stacking to ascertain the
average X-ray luminosity (in a given energy-band) produced due to
accretion onto the vast-majority of BHs, not only the BHs accreting
above a substantial fraction of Eddington, and subsequently ask how
this average accretion rate relates to the existence of a large-scale
bar.

Using {\sc stackfast}, we perform an X-ray stacking analysis
(described in Section~\ref{sec:stackfast}) on our spiral galaxy sample
based on their perceived bar properties. Given the known scaling
relations between the stellar content and the masses of BHs (e.g.,
\citealt{mcconnellma2013,Reines:2015aa}), we statistically negate the
effects of changing X-ray luminosities and Eddington ratios, due to
differing BH mass distributions, by stacking our parent sample in bins
complete in both stellar mass and redshift, i.e., the $M_* - z$
sub-samples that we describe in Section~\ref{sec:sample}. To also
ensure that no one stack is biased in sensitivity compared to the
other stacks, we used a K-S test to search for evidence of significant
differences in the distribution of exposure times for the sources
within the stacks. Based on the K-S statistic, we found no evidence
for significant differences in the $t_{\rm exp}$ distributions when
comparing the $f_{\rm bar}$ samples in a given $M_*-z$ bin.
Furthermore, to prevent bias towards artifically high stacked
luminosities due to significant individual detections, we do not
include the X-ray bright sources, and defer their analysis to
Section~\ref{subsec:brightsources}. In Figure~\ref{fig:lx_z}, we
present the results of the X-ray stacks performed in three energy
bands ($E=0.5$--2, 2--7, 0.5--7~keV) for our three $M_* - z$
sub-samples separated in bins of $f_{\rm bar}$.

Detailed {\it Chandra} studies of nearby star-forming and passive
galaxies have shown that in the absence of AGN activity, the X-ray
emission from hot diffuse gas, young high-mass X-ray binaries, and
older low-mass X-ray binaries correlates with the SFR and $M_*$,
respectively (e.g., \citealt{Gilfanov:2004aa,Revnivtsev:2007aa}). This
X-ray emission produced due to X-ray binaries ($L_{\rm X,SF}$) is an
obvious source of contamination in our total X-ray stacked luminosity
($L_{\rm X,tot}$). We estimate and statistically remove $L_{\rm X,SF}$
from $L_{\rm X,tot}$ by calculating the log-mean SFR and $M_*$ of the
stacked galaxy sample and invoking the local relations of
\cite{Lehmer:2010aa} and
\cite{Pereira-Santaella:2011aa}. Specifically, in the 0.5--2~keV band
we use Equation (4) presented in \cite{Pereira-Santaella:2011aa}, and
in the 2--7~keV band, we use the SFR-dependent relation
log~$L_{\rm X} = 39.46 + 0.76$log~SFR presented in
\cite{Lehmer:2010aa}, and linearly combine both $L_{\rm X,SF}$
predictions in the 0.5--7~keV. Both \cite{Lehmer:2010aa} and
\cite{Pereira-Santaella:2011aa} invoke a \cite{Kroupa:2001aa} initial
mass function to compute $M_*$ and SFRs in their models, this is
consistent with assumptions used for the SDSS $M_*$ and SFRs
measurements taken from \cite{brinchmann04}. After subtraction of
the average $L_{\rm X,SF}$ in the relevant bands, we produce clean
average X-ray luminosities due to BH accretion
($L_{\rm X, AGN} = L_{\rm X,tot} - L_{\rm X,SF}$).\footnote{We note
  that had we chosen to take the linear median of the SFRs and $M_*$,
  our average $L_{\rm X,SF}$ estimates would be systematically lower
  by 0.06~dex.} We observe average X-ray luminosities of
$L_{\rm X, AGN} \sim (0.1$--$2) \times 10^{40} \ergps$ for the stacked
samples. The stacks have typical S/N of $\sim 5$--10 calculated using
the stacked counts within the $r_{90}$ region, and the predicted
counts within $r_{90}$ using stacked local backgrounds (see Section
2.2. These results are presented in Table 1 for the three energy
ranges considered here.

In Figure~\ref{fig:lx_z} we show that within a given redshift bin
there is no significant difference (at the 90\% level) between the
average $L_{\rm X, AGN}$ for barred ($f_{\rm bar} > 0.25$), non-barred
($f_{\rm bar} < 0.1$), and ambiguously-barred
($f_{\rm bar} = 0.1-0.25$) spiral galaxies in SDSS-GZ2. We further
tested this result by selecting a sample of galaxies over the full
redshift range, $0.01<z<0.13$, within a relatively small stellar mass
range of $M_* \sim (3-10) \times 10^{10} M_{\odot}$, and performed an
additional stacking analysis. This sample included 58 barred, 36
ambiguously-barred, and 72 non-barred galaxies. We again found no
statistical difference between the average $L_{\rm X,AGN}$ in any of
the three energy ranges. Additionally, if $N_{\rm H}$ were playing a
significant role in our findings, then we would expect such a result
to manifest in the hardest energy band. However, we find no
significant difference between bars and non-barred galaxies in any of
the three X-ray energy ranges, suggesting that our result is
independent of obscuration effects.

While independent of the presence of a bar, we also find in
Figure~\ref{fig:lx_z} a trend of increasing $L_{\rm X,AGN}$ with
$z$. While the $f_{\rm bar}$ samples within a particular redshift bin
are matched in SFR and $M_*$, they are systematically different
between the redshift bins. Galaxies at higher-$z$ have systematically
higher $M_*$ in our sample due to the limiting magnitude of SDSS. Due
to the known relation between $M_*$ and $M_{\rm BH}$, a systematically
larger $M_*$ at a given $z$ will produce a higher average
$L_{\rm X,AGN}$ in each redshift bin.\footnote{Assuming a universal
  Eddington ratio distribution (e.g., \citealt{Aird:2012aa}), larger
  average $M_*$, and hence, larger average $M_{\rm BH}$, allows one to
  probe larger $L_X$. Higher $L_X$ that would be otherwise prohibited
  due to the Eddington limit for low-mass BHs, become feasible at
  larger $M_{\rm BH}$, i.e., the fraction of BHs capable of producing
  high $L_X$ becomes larger with increasing BH mass.} We show in the
upper panel of Figure~\ref{fig:lx_z} that when $L_{\rm X,AGN}$ is
normalized by average $M_*$ between the redshift bins, the resultant
specific accretion rate is independent of redshift in each energy
range. Thus, we can robustly conclude that there is no obvious
positive effect between the average X-ray luminosity associated with
BH accretion and the presence of a large-scale bar in nearby spiral
galaxies.

\subsection{Accretion onto X-ray detected AGN residing in barred spiral galaxies} \label{subsec:brightsources}

As part of our stacking analysis, we used forced photometry to
additionally detect ($S/N \gtrsim 5$) 101 (out of the 468 optical
galaxies) X-ray point-sources that were spatially coincident (median
offset of $0.26''$) with the optical nuclear position of the
galaxies. While each X-ray source is presumed to be an AGN, X-ray
emission from circumnuclear star-formation could still contribute to
the measured X-ray flux. Hence, we remove this contamination following
the same procedure used in the stacks. The X-ray detected AGN in our
sample have derived luminosities of
$L_{\rm X, 0.5-7keV} \sim 9 \times 10^{38}$--$5 \times 10^{43} \ergps$, similar
to X-ray luminous AGN identified in detailed studies of local
($D\lesssim 50$~Mpc) galaxies (e.g., \citealt{LaMassa:2011aa}).

We use the X-ray detected sources to further validate our stacking
results found in the previous section by producing an average stack
that includes both the detected and undetected X-ray sources. We
exclude the 16 AGN with $L_{\rm X,0.5-7keV} > 10^{42} \ergps$, which
form the higher luminosity tail of the AGN distribution in these
sources, and then stack all (X-ray detected and undetected) sources
observed with sufficient depth to detect an AGN if it were to have had
$L_X > 10^{42} \ergps$. This has the added effect of removing sources
from the stack with extremely shallow X-ray observations. With the
inclusion of the detected sources, we observed marginally higher
($\sim$0.17--0.26~dex) $L_{\rm X,AGN}$ in the stacked luminosities
when compared to the measurements presented in Table 1. However,
crucially, we still found no significant difference in the average
$L_{\rm X,AGN}$ between the $f_{\rm bar}$ sub-samples when including
the X-ray detected sources.

\begin{figure}[ht]
\centering
\includegraphics[width=0.99\linewidth]{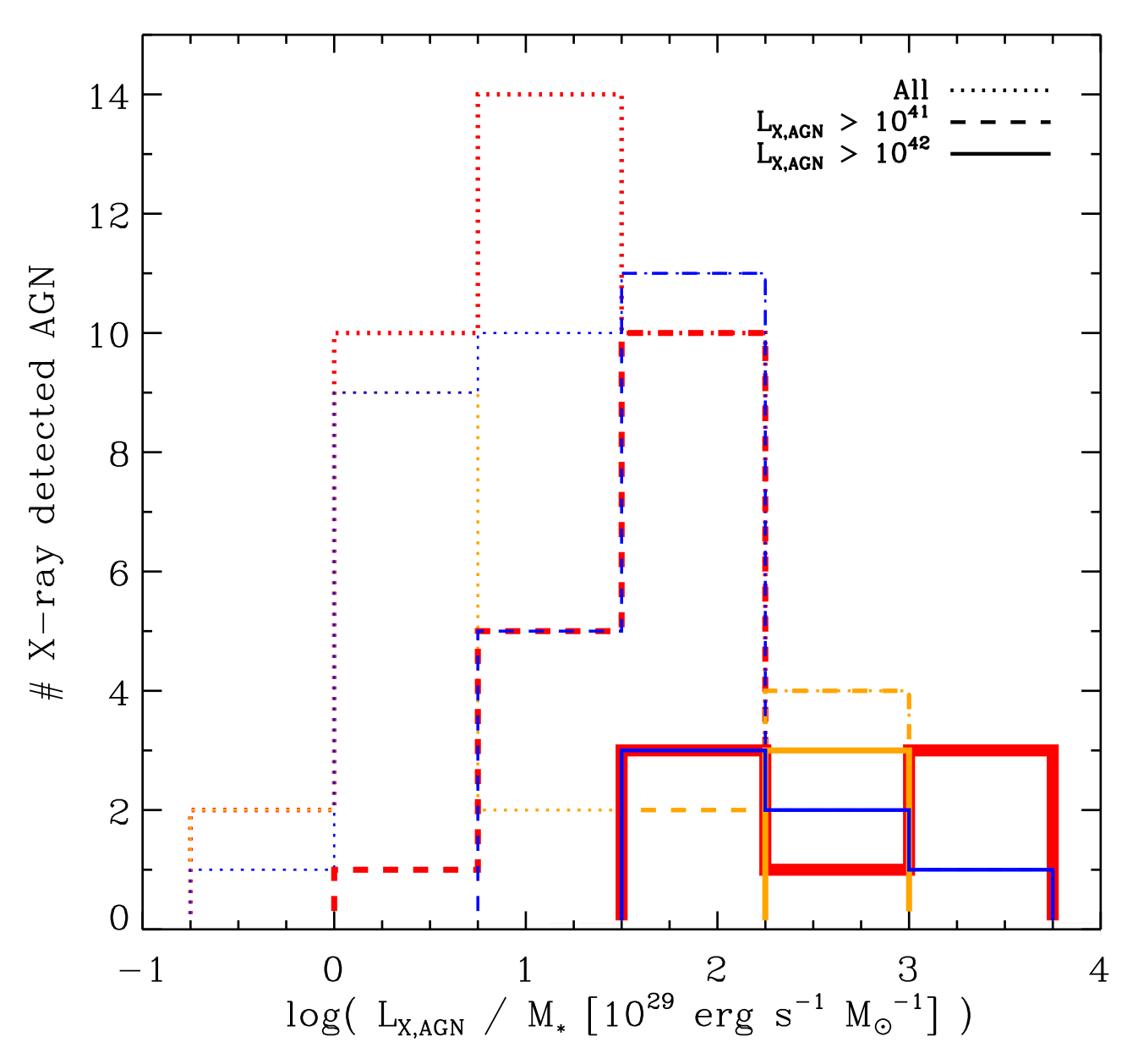}
\vspace{-0.1cm}
\caption{Specific accretion rate distributions ($L_{\rm X,0.5-7keV}/M_*$)
  for X-ray detected galaxies in our parent mass--SFR-matched SDSS-GZ2
  sample. Barred ($f_{\rm bar} > 0.25$), ambiguous
  ($0.1<f_{\rm bar}<0.25$) and unbarred ($f_{\rm bar} < 0.1$) spirals
  are plotted with red, orange and blue lines, respectively. Dotted,
  dashed and solid lines represent the full galaxy samples, with a cut
  of $L_X > 10^{41} \ergps$, and with a cut of $L_X > 10^{42} \ergps$,
  respectively.}
\label{fig:xdet}
\end{figure}

We find that the detected X-ray AGN luminosities are, in general, at
or above the stacked luminosities, suggesting that the stacks are
measuring the peak of the luminosity distribution of the overall
population, and the detected sources are populating a high-Eddington
tail of the accretion rate distribution. In Figure~\ref{fig:xdet} we
present the specific accretion rate ($L_X / M_*$) distributions for
the X-ray detected (at 0.5--7~keV) barred, ambiguous and unbarred
SDSS-GZ2 galaxies. Consistent with our finding of no dependence of
average accretion rate on the presence of a bar, we also show that
there is no difference between the specific accretion rates between
barred and non-barred galaxies for the X-ray detected AGN either
(two-sample KS reveals $P \sim 0.73$).

Differences between fueling of BHs due to the presence of a bar may
only become apparent when considering the most rapidly growing BHs in
the sample. Hence, we further investigated whether increasing
thresholds in X-ray luminosity, i.e., higher accretion rates, would
produce differences in the tails of the specific accretion rate
distributions for the galaxies in our sample. Applying cuts in X-ray
luminosity of $L_X > 10^{41} \ergps$ and $L_X > 10^{42} \ergps$, we
find that the peak of the distributions for barred and non-barred
galaxies shifts towards higher specific accretion rates. However, we
show in Figure~\ref{fig:xdet} that these peaks and the overall
distributions remain statistically the same between the barred and
non-barred systems, irrespective of the $L_X$ cuts. There is extremely
marginal evidence for an upturn at the highest specific accretion
rates probed by our sample
($L_X / M_* > 10^{32} \ergps M_{\odot}^{-1}$) for the barred galaxies,
though this is within the statistical uncertainties of the sample.

Taken together, we find no statistical evidence for a difference in
the specific accretion rate distributions of BHs present in barred or
non-barred galaxies. Thus, even over long $\sim$Gyr timescales, a
large-scale bar appears to have no effect on the growth of the central
BH in a spiral galaxy.

\section{Discussion \& Conclusions} \label{sec:summary}

We have investigated the X-ray luminosities of a large unbiased sample
of spiral galaxies selected from the SDSS-GZ2 based on the presence or
absence of a large-scale bar. Based on our X-ray stacking analysis of
archival {\it Chandra} X-ray data, we find that the average X-ray
luminosities of nearby ($z<0.15$) spiral galaxies are independent of
the existence of a bar. A similar result is also seen at
$z \sim 0.2$--0.8 where \cite{Cisternas:2015aa} observed no systematic
difference in the average X-ray luminosites between barred and
unbarred systems assessed from {\it HST} morphologies in the COSMOS
field. Indeed, allowing for the systematic shift to larger stellar
masses in the higher redshift sample, we find consistent X-ray
luminosities with the X-ray stacking analyses of
\cite{Cisternas:2015aa}. And consistent with results derived from more
samples of very nearby galaxies (e.g.,
\citealt{Ho:1997aa,Cisternas:2013aa}), we also find no observable
difference between the host galaxies of X-ray luminous AGN on the
basis of the presence of a bar.

In a given redshift bin, our stacked samples are matched in stellar
mass, and hence, on average, they can be considered to be also matched
in BH mass. Focusing on the $z\sim 0.05$--0.09 bin, where source
numbers are largest, we can use the local $M_* - M_{\rm BH}$ relation
\citep[see Equations (4) and (5) of][]{Reines:2015aa} to derive the
average accretion rates ($\dot{M}_{\rm acc}$) and $\lambda_{\rm Edd}$
for galaxies based on the presence (or absence) of a bar. The average
BH mass for the sample is $<$$M_{\rm
  BH}$$>$$\sim 2\times 10^7
M_{\odot}$.  Adopting a bolometric correction factor of $\kappa \sim
20$ \citep{vasudevan09b}, and assuming an average column density for
the sample of $N_{\rm
  H} = 3 \times 10^{21} \pcmsq$ and spectral slope
$\Gamma=1.8$,
the average bolometric luminosity, independent of the presence of a
bar is $L_{\rm
  bol,AGN} \approx 2 \times 10^{41}
\ergps$.  Further assuming a typical accretion efficiency of $\eta
\sim 0.1$, this corresponds to $\dot{M}_{\rm acc} \approx 3 \times
10^{-5}$~M$_{\odot}$~yr$^{-1}$, or $\lambda_{\rm Edd} \approx 2 \times
10^{-4}$. These low average accretion rates are typical of all very
nearby AGN, when sufficiently sensitive data is available to probe
these low accretion rates (e.g., \citealt{Goulding:2010aa}). Hence,
this suggests that, over long timescales, the presence of a bar has
little effect on the growth of BHs.

In order for our and previous analyses not to detect a significant
effect of a bar on the growth of the central BH, any causality between
these would need to be extremely short-lived, possibly occurring only
during the event that caused the bar to form (e.g., through a
minor-merger). Such an event could potentially provide a short-term
($\sim$100--300~Myrs)
supply of fuel for the BH, causing a rapid growth phase. However, the
resultant bar structure is substantially longer lived
($\sim$1--2~Gyrs;
\citealt{Bournaud:2005aa}), and will remain in-situ long after the BH
has exhausted any new fuel supply brought in through a
minor-merger. Our X-ray stacking analysis is particularly sensitive to
this long-lived phase, and shows that once the bar has formed, and
over its lifetime, these is no substantial net positive effect on the
growth of the BH.

\acknowledgments

We thank the anonymous referee for their careful reading and
insightful comments that significantly improved the manuscript. ADG
acknowledges support for this research from the National Aeronautics
and Space Administration through Chandra Award Number AR3-14016A
issued by the Chandra X-ray Observatory Center, which is operated by
the Smithsonian Astrophysical Observatory for and on behalf of the
National Aeronautics Space Administration under contract
NAS8-03060. DMA thanks the Science and Technology Facilities Council
(STFC) for support through grant ST/L00075X/1. The scientific results
reported in this article are based on data obtained from the Chandra
Data Archive. This research has made use of software provided by the
Chandra X-ray Center (CXC) in the application packages CIAO, ChIPS,
and Sherpa. This research also made use of the SDSS-III
database. Funding for SDSS-III has been provided by the Alfred
P. Sloan Foundation, the Participating Institutions, the National
Science Foundation, and the U.S. Department of Energy Office of
Science.

{\it Facilities:} \facility{Chandra (ACIS)}.

\bibliography{astroph_v1_0.bbl}

\end{document}